\documentclass{iau}
\usepackage{graphicx}

\title[IAUS291.~~Probing gravity with a pulsar-black hole system] 
{Prospects for probing strong gravity with a pulsar-black hole system} 

\author[N. Wex et al. ]  
{N.~Wex$^1$, 
K.~Liu$^{2,3}$, 
R.~P.~Eatough$^1$,
M.~Kramer$^{1,4}$, 
J.~M.~Cordes$^5$ \\ \and 
T.~J.~W.~Lazio$^6$
\thanks{Part of this research was carried out at the Jet 
Propulsion Laboratory, California Institute of Technology, under a contract with 
the National Aeronautics and Space Administration.}}

\affiliation{
$^1$Max-Planck-Institut f\"ur Radioastronomie, 
    Auf dem H\"ugel 69, D-53121 Bonn, Germany  \\ 
    email: {\tt wex@mpifr.de}
\\[\affilskip] 
$^2$Laboratoire de Physique et Chimie de l'Environnement,  LPCE UMR
6115 CNRS, \\
F-45071 Orleans Cedex 02, France
\\[\affilskip]   
$^3$Station de radioastronomie de Nancay, Observatoire de Paris, 
    CNRS/INSU, F-18330 Nancay, France
\\[\affilskip] 
$^4$University of Manchester, Jodrell Bank Centre for Astrophysics, 
    Alan-Turing Building, Manchester M13 9PL, UK
\\[\affilskip]    
$^5$Astronomy Dept., Cornell Univ., Ithaca, NY 14853, USA
\\[\affilskip] 
$^6$Jet Propulsion Laboratory, California Institute of Technology, 
    M/S 138-308, \\4800 Oak Grove Dr., Pasadena, CA 91109, USA
}

\pubyear{2012}
\volume{291}  
\jname{\mbox{Neutron Stars and Pulsars: Challenges and Opportunities after 80 years}}
\editors{J. van Leeuwen, ed.} 
\begin{document}

\maketitle

\begin{abstract}
The discovery of a pulsar (PSR) in orbit around a black hole (BH) is expected to 
provide a superb new probe of relativistic gravity and BH properties.  Apart from 
a precise mass measurement for the BH, one could expect a clean verification of 
the dragging of space-time caused by the BH spin. In order to measure the 
quadrupole moment of the BH for testing the no-hair theorem of general relativity 
(GR), one has to hope for 
a sufficiently massive BH. In this respect, a PSR orbiting the super-massive 
BH in the center of our Galaxy would be the ultimate laboratory for gravity tests 
with PSRs. But even for gravity theories that predict the same properties for 
BHs as GR, a PSR-BH system would constitute an excellent test 
system, due to the high grade of asymmetry in the strong field properties of 
these two components. Here we highlight some of the potential gravity 
tests that one could expect from different PSR-BH systems, utilizing present and 
future radio telescopes, like FAST and SKA.
\keywords{pulsars: general, black hole physics, relativity}
\end{abstract}




\noindent     
In the following we summarize the work presented in \cite{lwk+12}, \cite{liu12} 
and \cite{lewk13}. Based on these publications, we will highlight three different aspects of testing gravity with a PSR-BH system:
\begin{itemize}
\item Testing gravity with a PSR in orbit with a stellar mass 
      ($\sim 10 M_\odot$) BH. 
\item Testing gravity with a PSR in orbit around the BH in the Galactic 
      center, Sgr~A$^\ast$.
\item Testing scalar-tensor gravity with a PSR-BH system, as an example for the 
      probing power of a PSR-BH system for gravity theories, in which BHs are 
      the same as in GR.
\end{itemize}

The results given are based on extensive mock data simulations and consistent 
timing models, which are explained in detail in \cite{wk99}, \cite{lwk+12}, and 
\cite{liu12}. Furthermore, these simulations have been conducted for three different types of radio telescopes: a 100-m class telescope,
FAST and SKA. A detailed discussion on the expected timing precision for these
telescopes can be found in \cite{lwk+12} and \cite{liu12}. 


\section{Black hole properties}

One of the most intriguing results of general relativity (GR) is the uniqueness
theorem for the stationary BH solutions of the Einstein-Maxwell equations (see 
\cite{cch12}, and references therein). It implies that in GR all uncharged BH
solutions are described by the Kerr metric and, therefore, uniquely determined
by mass $M$ and spin $S$. Astrophysical BHs are believed to be the result of a
gravitational collapse, during which all the properties of the progenitor, apart 
from mass and spin, are radiated away by gravitational radiation (Price 
1972a,b\nocite{pri72a,pri72b}). The outer spacetime of an astrophysical BH should 
therefore (to a very good approximation) be described by the 
Kerr metric. Since the Kerr metric has a maximum spin at which it still exhibits 
an event horizon, Penrose's cosmic censorship conjecture (CCC) within GR 
(Penrose 1979\nocite{pen79}) requires the dimensionless spin parameter $\chi$ to satisfy
\begin{equation}
  \chi \; \equiv \; \frac{c}{G}\,\frac{S}{M^2} \; \le 1 \;
\end{equation}
A measured value for $\chi$ that exceeds 1 would pose a serious problem for our
understanding of spacetime, since this would indicate that either GR is wrong or
that a region may be visible to the outside universe, where our present
understanding of gravity and spacetime breaks down.

As a result of the no-hair theorem, all higher multipole moments ($l \ge 2$) of
the gravitational field of an astrophysical BH can be expressed as a
function of $M$ and $S$ (Hansen 1974\nocite{han74}). In particular, the 
dimensionless quadrupole moment $q$ satisfies the relation
\begin{equation}
  q \; \equiv \; \frac{c^4}{G^2}\,\frac{Q}{M^3} \; = \; -\chi^2 \;.
\end{equation}
A measurement of the quadrupole moment, in combination with a mass and a spin
measurement, would therefore identify the BH as a Kerr BH and provide a test of 
the no-hair theorem for BHs. 


\section{A pulsar in orbit with a stellar-mass black hole}

Concerning possible formation scenarios of PSR-BH systems, a detailed 
discussion is presented in \cite{liu12} and \cite{lewk13}. For the following,
it is only important to notice that there are several formation scenarios
that lead to a PSR-BH system with a recycled PSR.
 

\subsection{Mass determination}

\begin{figure}[ht]
\begin{center}
 \includegraphics[width=3.3in]{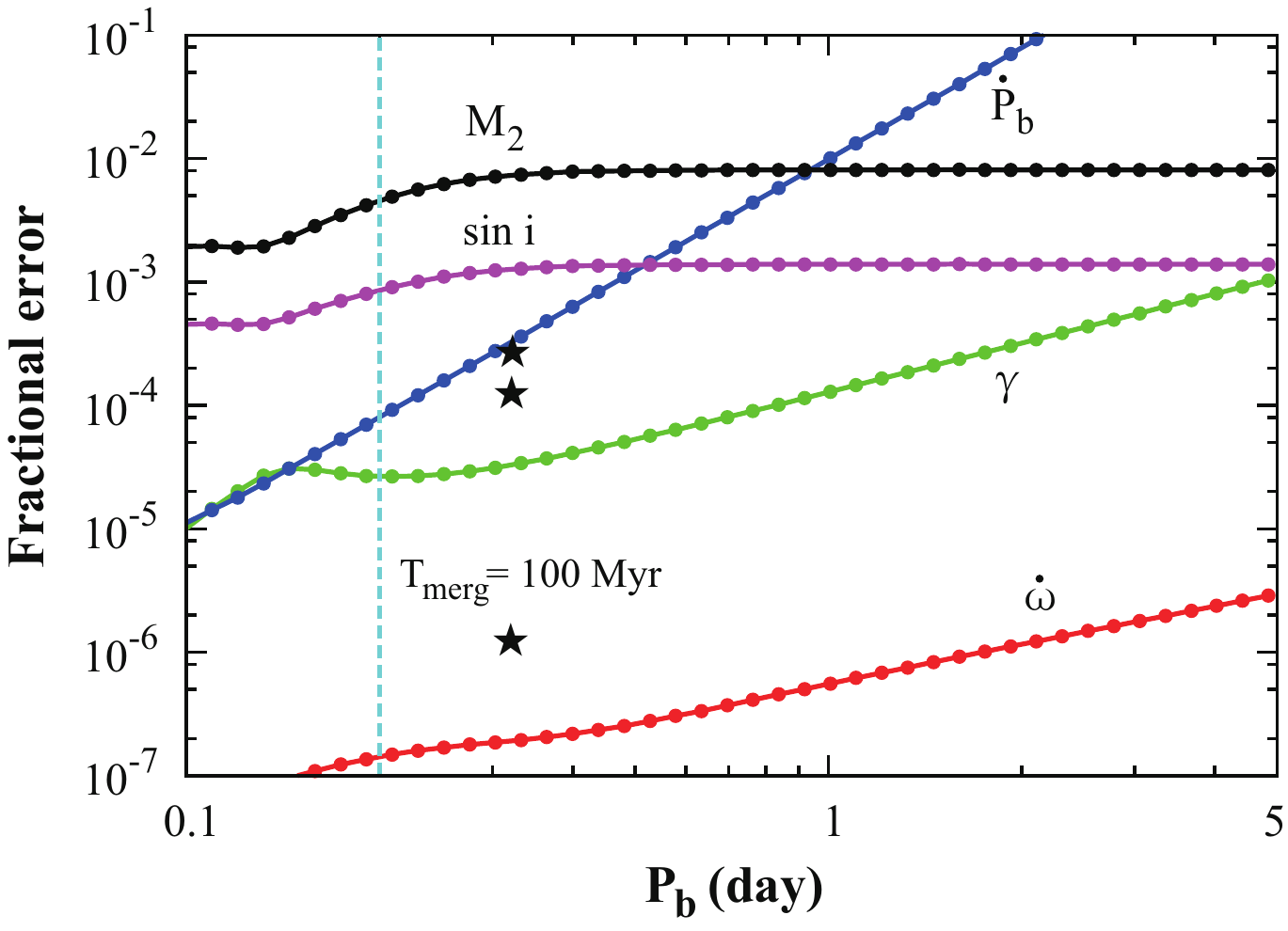} 
 \caption{Fractional error for different PK parameters, which can be used to 
 determine the mass of the BH, as a function of the orbital period. The 
 figure is based on mock data simulations for weekly observations over 5
 years with a 100-m 
 class telescope, assuming a recycled PSR in a mildly eccentric ($e = 0.1$) 
 orbit with a 10\,$M_\odot$ BH. Note that $\dot\omega$ is expected to 
 have a significant contribution from the frame-dragging caused by the rotation
 of the BH, and can {\it a priori} not be used in the mass determination. 
 For a comparison, the 
 three stars mark the precision obtained in the Hulse-Taylor pulsar (top to bottom: $\dot{P}_b$, $\gamma$, $\dot\omega$; \cite{wnt10}).} 
\label{fig:bhm}
\end{center}
\end{figure}

The most precise mass measurements for stars (other than the Sun) come from 
PSR timing observations (Lorimer \& Kramer 2005, Weisberg \etal\ 
2010\nocite{lk05,wnt10}). 
Those are achieved in binary 
PSR systems, where in addition to the Keplerian parameters one can determine a 
set of post-Keplerian (PK) parameters that describe the relativistic correction 
in the orbital motion and signal propagation. In GR (and many alternative 
theories of gravity) the PK parameters are functions of the Keplerian parameters 
and the two {\it a priori} unknown masses of the system, which can be determined 
once two PK parameters have been obtained (Lorimer \& Kramer 2005\nocite{lk05}). 
In a PSR-BH system the observation of PK parameters will allow the determination 
of the BH mass with unprecedented precision. 
Fig.~\ref{fig:bhm} (\cite[taken from  Liu et al., in prep.]{lewk13}) illustrates the 
precision that one can expect with a 100-m class telescope. Future telescopes 
like FAST and SKA would yield a significantly higher precision, as demonstrated by the simulations in \cite{liu12} and \cite{lewk13}.

\subsection{Frame dragging}
\label{sec:bh_spin}

\begin{figure}[ht]
\begin{center}
\includegraphics[width=3.3in]{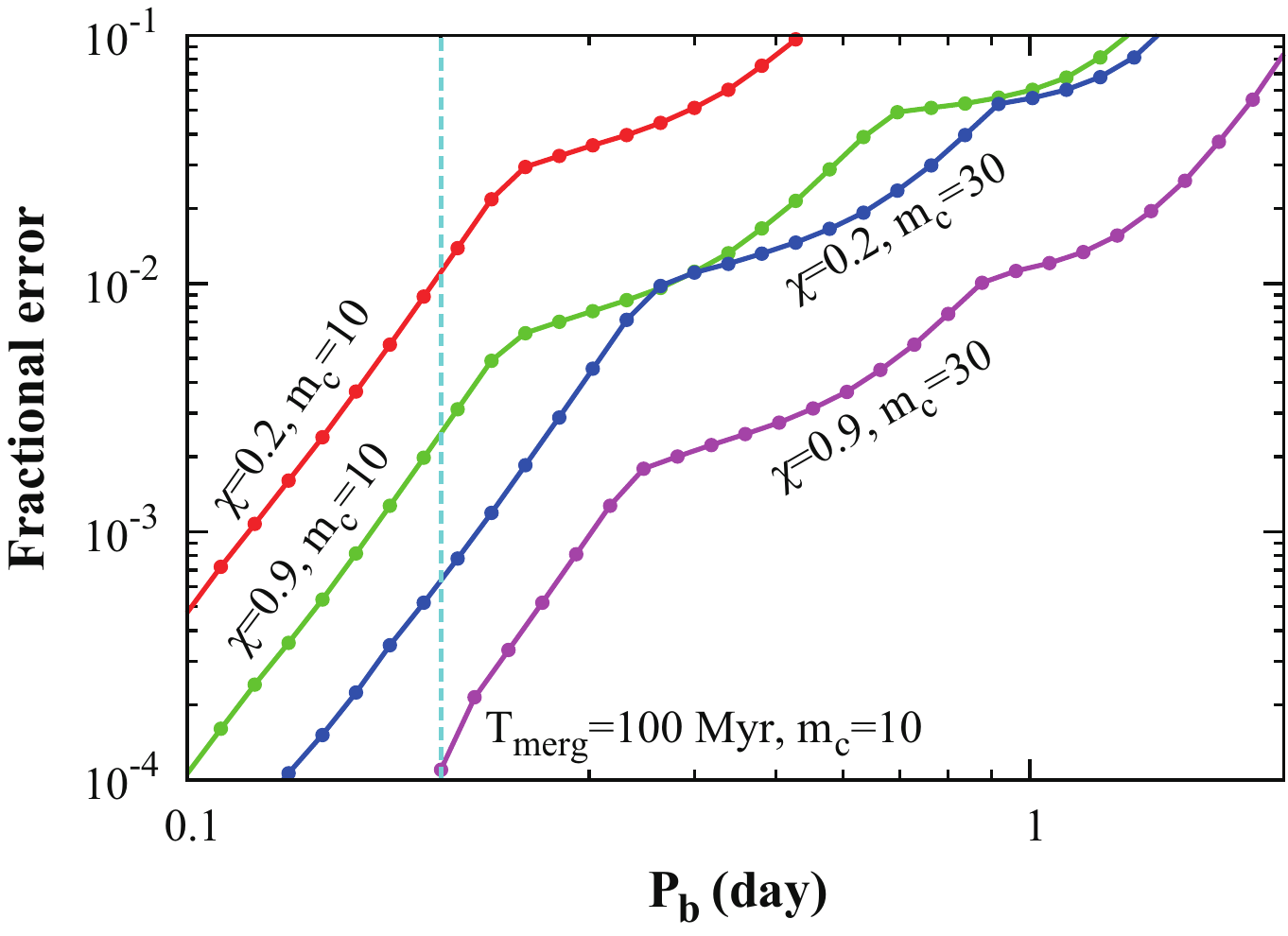} 
\caption{Fractional error in the BH spin, as a function of the orbital period. 
 The figure is based on mock data simulations for weekly observations over
 5 years with SKA, 
 assuming a recycled PSR in a mildly eccentric ($e = 0.1$) orbit. $m_c$ denotes
 the BH mass. 
\label{fig:spin}}
\end{center}
\end{figure}

It has been shown by \cite{wk99} that the Lense-Thirring precession of the orbit,
caused by the relativistic spin-orbit coupling, between the orbital angular 
momentum and 
the BH spin, is the best way to measure the magnitude and direction of the BH 
spin. Once the second time derivatives of the longitude of periastron and 
the projected semi-major axis become observable in the timing data,
the spin can be determined. Fig.~\ref{fig:spin} (\cite[taken from  Liu et al., in prep.]{lewk13}) illustrates the precision that one can expect with SKA.
A spin measurement in a PSR-BH system would verify the frame dragging 
caused by a rotating BH, and test the CCC inequality $\chi \le 1$.

\subsection{Quadrupole moment}

In order to perform a test of the no-hair theorem, it would be necessary to 
measure the quadrupole moment of the BH. As already argued by \cite{wk99},
such a measurement has to be based on the periodic features in the orbital 
motion, caused by the unique structure of the gravitational potential 
associated with the quadrupole moment. Our simulations, presented in 
\cite{liu12} and \cite{lewk13} show, 
that only for a very massive BH ($\gtrsim 30\,M_\odot$) and very tight orbits 
($P_b \lesssim 0.2\,{\rm d})$ one can hope to measure the quadrupole signature in 
the data, provided one has the superb timing precision of the SKA. We consider it 
as not very likely, that such a system can be found in nature, in particular 
since such a system would have a very short lifetime in the order of a few Myr.


\section{A pulsar in orbit around Sgr A$^\ast$}

According to the previous section, a system consisting of a radio PSR and 
a stellar mass BH will most likely not allow the determination of 
the BH quadrupole moment to test the no-hair theorem. Since the quadrupole 
moment becomes more important for more massive BHs, one wishes to find a PSR in 
orbit around the most massive BH in our 
Galaxy, Sgr~A$^\ast$ in the Galactic center ($M \sim 4 \times 10^6\,M_\odot$; 
\cite[Ghez \etal\ 2008, Gillessen \etal\ 2009]{gsw+08,get+09}). Finding and timing a PSR
in orbit around Sgr~A$^\ast$ comes with certain challenges, which are discussed 
in more details in \cite{lwk+12}, Eatough \etal\ (this proceedings) and 
references therein. In short, it is more
likely to detect a young PSR with a rotational period $\sim$ 0.5\,s, which 
then would give us a timing precision in the order of $\sim$ 0.1 to 1\,ms. All this requires observational frequencies $\gtrsim 20$\,GHz. Furthermore, one 
would also have to worry about perturbations by other masses in the immediate
vicinity of Sgr~A$^\ast$ (see \cite[Liu \etal\ 2012]{lwk+12}, and references therein). In the
results presented here we assume a clean orbit and a  
weekly time-of-arrival measurement with 0.1\,ms uncertainty. As we will see, this
is already sufficient, to go all the way to the no-hair theorem test, i.e.\ 
measuring the mass $M$, spin $S$ and quadrupole moment $Q$ of Sgr~A$^\ast$ with 
sufficient precision.


\subsection{Mass determination and the distance to the Galactic center}

For the precision that can be obtained in determining the mass of Sgr~A$^\ast$ 
from PK parameters, the PSR can be seen as a test mass in orbit 
around Sgr~A$^\ast$. In this case the measurement of one PK parameter is 
sufficient. A fractional precision of $10^{-5}$, or even better, should be easily 
possible with a PSR in a $\lesssim$ 1\,yr orbit (Liu \etal\ 2012\nocite{lwk+12}). 
Such a precision is not only key 
to extract the Lense-Thirring contribution to the peri-center precession 
$\dot\omega$, it would also allow, in combination with astrometric observations 
of the S-stars, to determine the distance to the Galactic center $R_0$, since, 
unlike the astrometric mass, the mass measurement from PSR timing is not affected 
by an uncertainty in $R_0$.\footnote{According to our estimates, in combination 
with a $10\,\mu$as infrared astrometry, an uncertainty of just a few pc could be 
reached.}  


\subsection{Frame dragging}

Although there is clear indication that Sgr~A$^\ast$ rotates, its actual rate of 
rotation is still not well determined, and a rather large range in the estimates 
of $\chi$ can be found in the literature (see references in 
\cite[Liu \etal\ 2012]{lwk+12}), 
which is a result of the uncertainty in the underlying model assumptions. A 
PSR would, in the absence of any major external perturbations, give direct 
access to the dragging of inertial frames in the vicinity of Sgr~A$^\ast$. Like 
in the case of stellar mass BHs, the (additional) precession of a PSR 
orbit due to the frame dragging (Lense--Thirring precession) is the most promising 
effect to determine the direction 
and magnitude of the BH spin. \cite{lwk+12} have used extensive mock
data simulations to show, that for orbits below an orbital period of 0.5 years, 
where external perturbations are likely to be negligible, the spin parameter 
$\chi$ could be measured with a precision of $10^{-4}$ to $10^{-3}$, or even 
better in case of very short orbital periods ($P_b \lesssim 0.1$\,yr) and/or 
high eccentricities ($e \gtrsim 0.8$). This would be a test of the frame dragging 
caused by a super-massive BH and, like in section~\ref{sec:bh_spin}, a test of 
the CCC inequality $\chi \le 1$.


\subsection{No-hair theorem test}

For a BH, with a given spin parameter $\chi$, the quadrupole moment grows 
very fast with increasing mass ($\propto M^3)$. For this reason, even for rather 
wide orbits of several 10\,AU the quadrupole moment of Sgr~A$^\ast$ is expected 
to give rise to measurable effects in the orbital motion. The calculations in 
\cite{lwk+12} clearly show, that the observable amplitudes of the ``quadrupole 
effect'', for an orbit of a few 0.1\,yr, are of order of several milli-seconds, 
and should allow a better than 1\,\% test of the no-hair theorem. In contrast to
an astrometric no-hair theorem test (Will 2008\nocite{will08}), only one star (here the PSR) is needed and the orbital eccentricity can be considerably less extreme (see Fig.~10 in \cite[Liu \etal\ 2012]{lwk+12}).


\section{Scalar-tensor gravity and PSR-BH systems}

\begin{figure}[ht]
\begin{center}
\includegraphics[width=4.35in]{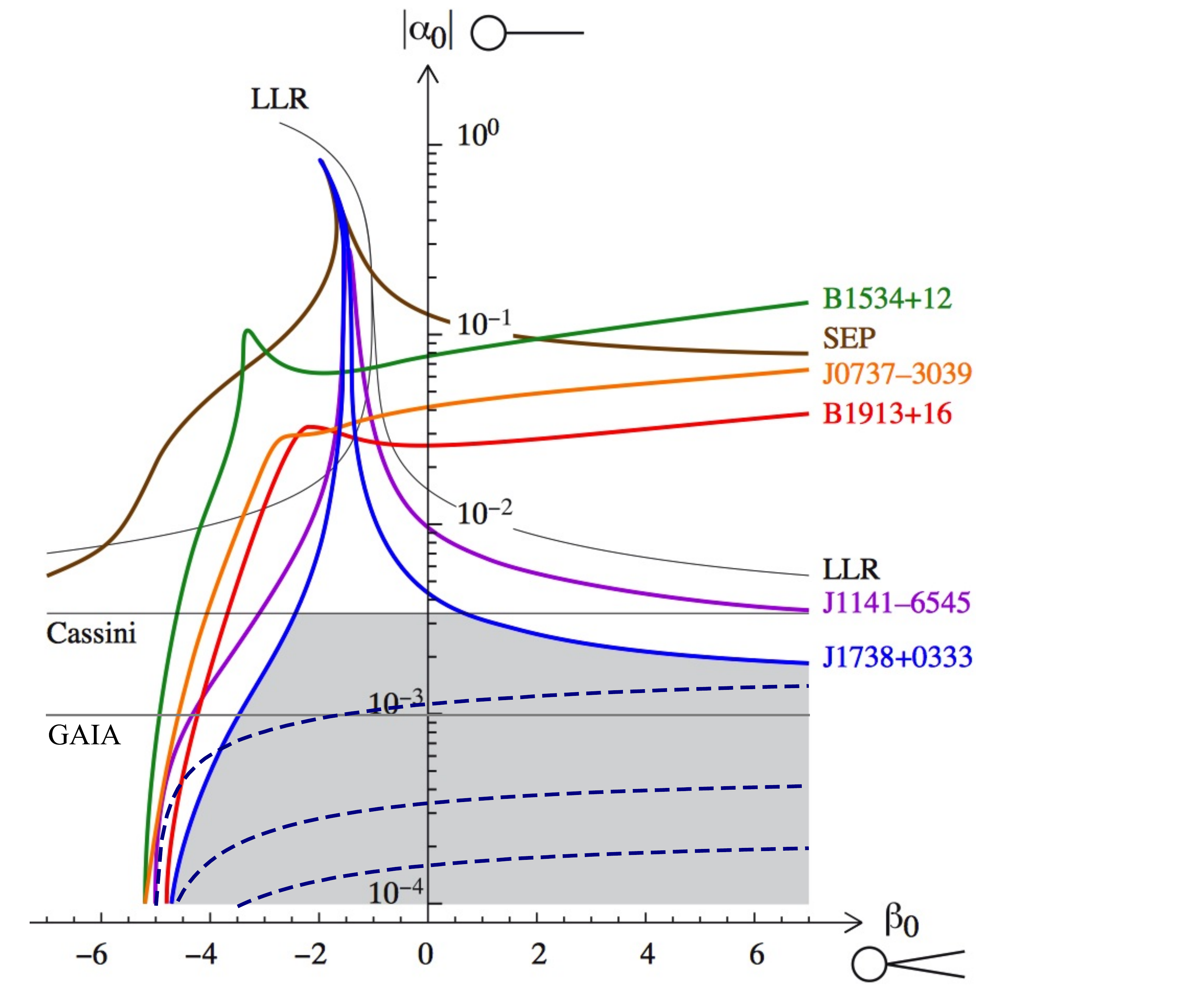} 
\caption{Constraints in the $\beta_0$--$\alpha_0$ plane (different PSRs 
and a hypothetical PSR-BH system). Details are given in \cite{fwe+12} and 
\cite{lewk13}. Limits by a hypothetical PSR-BH system, consisting 
of a recycled PSR in orbit with a 10\,$M_\odot$ BH ($P_b = 5\,{\rm d}$, 
$e = 0.8$) are given by dashed lines, which exclude the region above and left of 
them. The three lines, correspond to (top to bottom): 10\,yr weekly timing with a 
100-m class telescope, 5\,yr weekly timing with FAST, and 5\,yr weekly timing 
with SKA.
\label{fig:st}}
\end{center}
\end{figure}

In the previous sections we have seen the potential to probe the properties of 
gravity with a PSR in orbit around a BH. But even for gravity theories that 
predict the same properties for a BH as GR, a PSR-BH system would be an excellent 
probe, that can be superior to all present binary PSR experiments. An 
important class of gravity theories where this is the case, as pointed out by 
\cite{de98}, are scalar-tensor theories. Scalar-tensor gravity, in which gravity 
is mediated by a tensor field $g_{\mu\nu}^\ast$ and by a massless scalar 
field $\varphi$, are well motivated and physically consistent alternatives to GR, 
that have been studied extensively (see \cite{fm03}). In the scalar-tensor 
gravity, investigated in detail in \cite{de96}, the orbital motion of a binary 
system depends, besides the Einstein masses $m_A$, $m_B$ on the effective 
coupling constants $\alpha_a = \partial \ln m_a / \partial\varphi_0$ ($a=A,B$), 
and their scalar-field derivatives $\partial \beta_a / \partial\varphi_0$, where 
$\varphi_0$ denotes the asymptotic value of $\varphi$ at spatial infinity. In a
PSR-BH system, we have $\alpha_{\rm BH} = 0$ (and consequently 
$\beta_{\rm BH} = 0$) because of the no-scalar-hair theorems, which considerably 
simplifies the equations for the PK parameters given in \cite{de96}. Of 
particular interest here is the leading term in the gravitational wave damping 
(dipolar radiation) of the orbital period $P_b$, that is simply
\begin{equation}
  \dot{P}_b^{\rm dipolar} \simeq -\frac{4\pi^2G}{c^3 P_b}\,
  \frac{m_{\rm PSR}m_{\rm BH}}{m_{\rm PSR} + m_{\rm BH}} \,
  \frac{1 + e^2/2}{(1 - e^2)^{5/2}} \,
  \alpha_{\rm PSR}^2\;.
\end{equation}
For a given equation-of-state, the effective scalar coupling of the PSR, 
$\alpha_{\rm PSR}$, depends on the two fundamental coupling parameters
of the theory, $\alpha_0$ (linear) and $\beta_0$ (quadratic), and
the mass of the PSR, which we assume to be the canonical value of 
1.4\,$M_\odot$. \cite{lewk13} have conducted extensive mock data 
simulations for different types of telescopes and different orbital parameters  
to demonstrate the superb capabilities of a PSR-BH binary to constrain 
scalar-tensor gravity. Figure~\ref{fig:st} presents some of these results. As
can be seen there, in particular with future radio telescopes 
a PSR-BH system would provide an extremely sensitive test for the presence of 
a scalar degree of freedom in the gravitational interaction.
 
\end{document}